%% file: main.tex
\documentclass[conference]{IEEEtran}
\IEEEoverridecommandlockouts
\input{packages}
\def\BibTeX{{\rm B\kern-.05em{\sc i\kern-.025em b}\kern-.08em
    T\kern-.1667em\lower.7ex\hbox{E}\kern-.125emX}}
\newcommand{\junk}[1] {}
\input{notation_commands}
\renewcommand{\baselinestretch}{0.945} 

\begin{document}

\title{Learning without Data: Physics-Informed \\Neural Networks for Fast Time-Domain Simulation\\

\thanks{This work is supported by the multiDC project funded by Innovation Fund Denmark, Grant No. 6154-00020B, and by the ERC Project VeriPhIED, funded by the European Research Council, Grant Agreement No: 949899}
}

\author{\IEEEauthorblockN{Jochen Stiasny, Samuel Chevalier, and Spyros Chatzivasileiadis}
\IEEEauthorblockA{Department of Electrical Engineering\\
Technical University of Denmark\\
Kgs. Lyngby, Denmark \\
\{jbest,schev,spchatz\}@elektro.dtu.dk}
}

\maketitle

\input{sections/00_abstract}

\begin{IEEEkeywords}
Runge-Kutta, neural networks, time-domain simulation, transient stability analysis
\end{IEEEkeywords}

\input{sections/01_introduction}
\input{sections/02_methodology}
\input{sections/03_case_study}
\input{sections/04_results}

\input{sections/05_conclusion}

\bibliographystyle{ieeetr}
\bibliography{RK_bib}

\end{document}

%% file: packages.tex
\usepackage{amsmath}
\usepackage{amsfonts}
\usepackage{amssymb}

\usepackage{cite}
\usepackage{algorithmic}
\usepackage{graphicx}
\usepackage{textcomp}
\usepackage{booktabs}
\usepackage{xcolor}
\usepackage{bm}
\usepackage[capitalise]{cleveref}
\usepackage{siunitx}

\usepackage{tikz,pgf} 
\usepackage{lipsum}
\usetikzlibrary{fit}
\usepackage{pgfplotstable}
\pgfplotsset{compat = 1.15}
\usepgfplotslibrary{statistics}
\usetikzlibrary{intersections}
\usepgfplotslibrary{fillbetween}
\usepgfplotslibrary{colorbrewer}
\usetikzlibrary{pgfplots.statistics, pgfplots.colorbrewer} 
\usepgfplotslibrary{statistics}
\pgfplotsset{cycle list/OrRd-3}
\usepgfplotslibrary{groupplots}
\usetikzlibrary{plotmarks}
\usetikzlibrary{decorations.pathreplacing}
\usetikzlibrary{spy}

\DeclareSIUnit\pu{p.u.}

%% file: notation_commands.tex
\newcommand{\timestep}{\ensuremath{\Delta t}}
\newcommand{\RKalpha}[1]{\ensuremath{\alpha^{#1}}}
\newcommand{\RKbeta}[1]{\ensuremath{\beta^{#1}}}
\newcommand{\RKgamma}[1]{\ensuremath{\gamma^{#1}}}

\newcommand{\xinitial}{\ensuremath{\bm{x}^0}}
\newcommand{\tinitial}{\ensuremath{t_0}}
\newcommand{\xprediction}{\ensuremath{\bm{x}^1}}
\newcommand{\xpredictionhat}{\ensuremath{\hat{\bm{x}}^1}}

\newcommand{\RKstage}[1]{\ensuremath{\bm{h}^{#1}}}

\newcommand{\RKstagehat}[1]{\ensuremath{\hat{\bm{h}}^{#1}}}

\newcommand{\RKerror}[1]{\ensuremath{\bm{\epsilon}^{#1}}}
\newcommand{\RKerrori}[1]{\ensuremath{\epsilon^{#1}_i}}

\newcommand{\deltaprediction}{\ensuremath{\delta^1}}
\newcommand{\deltapredictionhat}{\ensuremath{\hat{\delta}^1}}
\newcommand{\omegapredictionhat}{\ensuremath{\hat{\omega}^1}}

\newcommand{\deltaerror}{\ensuremath{e_{\delta}}}

\newcommand{\physicsError}{\ensuremath{\bm{\xi}}}

\newcommand{\lossGeneral}{\ensuremath{\mathcal{L}}}
\newcommand{\lossRKstage}{\ensuremath{\mathcal{L}_i^k}}
\newcommand{\weightRKstage}{\ensuremath{\lambda_i^k}}
\newcommand{\lossPredictionDiff}{\ensuremath{\mathcal{L}_i^{dt}}}
\newcommand{\weightPredictionDiff}{\ensuremath{\lambda_i^{dt}}}

\newcommand{\fvector}{\ensuremath{\bm{f}}}
\newcommand{\statevariable}{\ensuremath{\bm{x}}}
\newcommand{\inputvariable}{\ensuremath{\bm{u}}}

\newcommand{\weights}[1]{\ensuremath{\bm{W}_{#1}}}
\newcommand{\biases}[1]{\ensuremath{\bm{b}_{#1}}}
\newcommand{\perceptron}[1]{\ensuremath{\bm{z}_{#1}}}
\newcommand{\outputNN}{\ensuremath{\bm{y}}}

%% file: sections/00_abstract.tex
\begin{abstract}
In order to drastically reduce the heavy computational burden associated with time-domain simulations, this paper introduces
a Physics-Informed Neural Network (PINN) to directly learn the solutions of power system dynamics. In contrast to the limitations of classical model order reduction approaches, commonly used to accelerate time-domain simulations, PINNs can universally approximate any continuous function with an arbitrary degree of accuracy. One of the novelties of this paper is that we avoid the need for any training data. We achieve this by incorporating the governing differential equations and an implicit Runge-Kutta (RK) integration scheme directly into the training process of the PINN; through this approach, PINNs can predict the trajectory of a dynamical power system at any discrete time step. The resulting Runge-Kutta-based physics-informed neural networks (RK-PINNs) can yield up to 100 times faster evaluations of the dynamics compared to standard time-domain simulations. We demonstrate the methodology on a single-machine infinite bus system governed by the swing equation. We show that RK-PINNs can accurately and quickly predict the solution trajectories.
\end{abstract}

%% file: sections/01_introduction.tex
\section{Introduction}\label{sec:introduction}
Time-domain simulation is a fundamental tool for ensuring operational stability of modern electrical power systems. Over the last many decades, power system engineers have developed a mature catalogue of physics-based modeling strategies, e.g.,~\cite{Kundur:1994,Sauer:2006}, and numerically efficient simulation platforms, e.g.,~\cite{Milano:2005,Long:1990}. Building on these advances, emerging synthetic test case libraries~\cite{Xu:2018,Xu:2017} are allowing researchers to explore the dynamical complexity of massive power system models which contain tens of thousands of buses.

As grid dynamics become both faster and more distributed, however, the computational expense associated with running full-scale power system simulations is growing considerably. Traditionally, power system engineers have relied on various Model Order Reduction (MOR) methodologies~\cite{Chaniotis:2005,Ramirez:2016,Zhang:2018} and modal analysis techniques~\cite{Crow:2005,Trudnowski:1999} to overcome the computational burden and complexity of large models. Accordingly, dynamic model reduction tools are built into a variety of commercial simulators (e.g., DIgSILENT, DYNRED and PSS/E)~\cite{Milano:2009} and are commonly used by industry.

While such equivalency tools can produce compact surrogate models, standard MOR has significant drawbacks in certain applications. For example, the majority of MOR techniques used in power systems are only applicable for linear systems (e.g., prony analysis and matrix pencil methods~\cite{Trudnowski:1999,Crow:2005}; Autoregressive models~\cite{Chai:2016}; Gramiam-based approaches~\cite{Ramirez:2016}). Second, most MOR methods are projection based, and the resulting dynamical model must still be directly simulated by an ODE solver. Finally, most classical MOR tools which are applicable to power systems cannot efficiently compress any given nonlinear function with an arbitrary degree of accuracy. For example, the Koopman Mode Decomposition requires infinite terms to approximate a logistic map~\cite{Brunton:2019}, and Sparse Identification of Nonlinear Dynamics (SINDY)~\cite{Loiseau:2018} may fail if the nonlinear basis functions are poorly chosen.

In contrast to that, Neural Networks (NNs) have the capacity to universally approximate any continuous function with an arbitrary degree of accuracy~\cite{Leshno:1993}. Therefore, they are able to overcome all of the aforementioned drawbacks, and they have become a popular alternative to classical MOR approaches. Recently, the so-called Physics-Informed Neural Network (PINN)~\cite{Raissi:2019} was proposed as a framework for directly mapping dynamical system inputs (initial conditions) to system outputs (state trajectories). PINNs naturally allow for a direct regularisation of the training process with physical sensitivities, and they also bypass the need for a numerical solver altogether, thus approximating the solution of the underlying ODE system. In~\cite{Raissi:2019}, continuous and discrete time PINN approaches were presented. In the discrete time formulation, the NN predicts the numerical values associated with an implicit Runge Kutta (RK) integration scheme. The training process of this NN, subsequently referred to as RK-PINNs, does not require any simulated data; hence, it naturally yields a ``data free" framework.

Physics-informed neural networks have been first introduced in power systems in our previous work~\cite{misyris_physics-informed_2020}, and, since then, have extended in applications related to system identification~\cite{stiasny2021physicsinformed}, the transient response of interconnected systems~\cite{stiasny_transient_2021}, and DC optimal power flow \cite{Nellikkath:2021}; along the same lines, sensitivity-informed NNs have been recently introduced for AC power flow optimization~\cite{Singh:2021} and physics-informed graphical NNs for parameter estimation~\cite{Pagnier:2021}. All of these works, however, have utilised a continuous formulation of the problem's underlying physics and have thus required simulated training data.

Given (i) the increasing computational complexity of time-domain integration approaches, (ii) the potential inadequacy of classical MOR tools to approximate all newly emerging non-linearities, e.g., related to converter-dominated systems, and (iii) the emerging success of PINN modeling for power systems, this paper leverages the RK-PINNs framework in order to learn power system dynamics with a ML model. The resulting model can be used, e.g., to screen transient stability contingencies orders of magnitude faster than conventional numerical integration schemes, such as Runge-Kutta, and most attractively, the RK-PINN model parameters can be learned without the need for any simulated power system training data. Accordingly, our contribution lies in
\begin{itemize}
    \item introducing, for the first time, the RK-PINNs learning framework in power systems applications;
    \item extending the fundamental structure of RK-PINNs to incorporate variable time steps;
    \item introducing an additional regularisation term on the RK-PINN's prediction based on the differential equations;
    \item and providing a publicly available code base.
\end{itemize}

After introducing the methodology in \cref{sec:methodology}, we describe the case study in \cref{sec:case_study} and the results in \cref{sec:results}. We conclude in \cref{sec:conclusion}.



%% file: sections/02_methodology.tex
\section{Methodology}\label{sec:methodology}

We first describe the two fundamental elements of the method, i.e., implicit Runge-Kutta (RK) integration schemes and neural networks. Then, we show how neural networks can assist in solving the non-linear system of equations stemming from the RK scheme. As the methodology is not limited to a specific problem, we introduce it in a general form in this section, and then we show its application to power systems in \cref{sec:case_study}. 

\subsection{Runge-Kutta schemes}

RK methods allow us to find an approximation of the temporal evolution of a dynamical system described by a set of ordinary differential equations (ODEs)
\begin{align}
    \frac{d}{dt}\statevariable &= \fvector(t, \statevariable(t); \inputvariable). \label{eq:state_equation}
\end{align}
In \eqref{eq:state_equation}, \statevariable{} represents the state vector that evolves over time $t$ and \inputvariable{} represents control inputs to the system. Function \fvector{} describes the parametrised update rule for \statevariable{}. These systems of ODEs can be solved by the following general RK scheme~\cite{Raissi:2019} for a time step \timestep{}:
\begin{align}
    \RKstage{k} &= \fvector \left(\tinitial + \RKgamma{k}\timestep, \xinitial + \timestep \sum_{l=1}^s \RKalpha{kl} \RKstage{l} ; \inputvariable \right), \; k = 1, \dots, s, \label{eq:RK_scheme}\\
    \xprediction &= \xinitial + \timestep \sum_{k=1}^s \RKbeta{k} \RKstage{k}. \label{eq:RK_scheme_prediction}
\end{align}
The $s$ RK-stages \RKstage{k} represent state update vectors according to \eqref{eq:RK_scheme}. The prediction \xprediction{} of the system state at time $\tinitial +\timestep{}$ can afterwards be calculated by weighing these state update vectors \RKstage{k} and applying it for a \timestep{} onto the initial condition \xinitial{} as shown in \eqref{eq:RK_scheme_prediction}. The properties of the integration scheme largely depend on the coefficients \RKalpha{kl}, \RKbeta{k}, and \RKgamma{k}. Their values determine whether the scheme is explicit or implicit and the order of the scheme which governs the truncation error. If the coefficient matrix \RKalpha{kl} has a strictly lower-triangular shape, the scheme is explicit; otherwise, it contains implicit functions of \RKstage{k}. Typical RK-schemes include the forward and backward Euler, the common RK-45 scheme, the trapezoidal method, and many more regularly used schemes. Implicit schemes find use mostly in stiff ODEs as explicit schemes would require a very small time step size due their numerical stability properties. However, the resulting system of non-linear equations can be difficult to solve, in particular for large number of stages $s$ and systems with many states.

\subsection{Neural networks}

The reason to use neural networks (NNs) for this problem boils down to exploiting their capacity to approximate the solution to the system of non-linear equations that \eqref{eq:RK_scheme} yields with an arbitrary degree of accuracy, while providing extremely quick evaluations. The NN training problem takes the form
\begin{alignat}{2}
    \min_{\weights{i}, \biases{i}} \quad & \lossGeneral && \label{eq:NN_loss_term}\\
     \perceptron{k+1} &= \sigma(\weights{k+1} \perceptron{k} + \biases{k+1}),\; && \forall k = 0, 1, ..., K-1\label{eq:NN_hidden_layers}\\
    \outputNN &= \weights{K+1} \perceptron{K} + \biases{K+1}.&&\label{eq:NN_output}
\end{alignat}
\weights{i} and \biases{i} represent the adjustable parameters, namely the weight matrix and bias vector of the $i$-th layer. Equation~\eqref{eq:NN_hidden_layers} describes the hidden layers of the NN in which the input to the layer \perceptron{i} undergoes a linear transformation followed by an element-wise non-linearity, e.g., $\tanh$, to yield the output of the layer \perceptron{i+1}. The neural network consists of $K$ hidden layers. \perceptron{0} equals the input vector and \eqref{eq:NN_output} describes the final layer in which we only apply a linear transformation to obtain the NN's ouput \outputNN{}. We will address the formulation of  \perceptron{0}, \outputNN{}, and the objective \lossGeneral{} in \eqref{eq:NN_loss_term} in the following subsections where we incorporate an implicit RK-scheme into this general NN. In this context, we will make use of the tool of automatic differentiation (AD) that allows us to evaluate the derivative of the output variables with respect to the NN's inputs.

\subsection{Fixed time step RK-PINN}

In a first step to incorporate the RK-scheme into NNs, similar to \cite{Raissi:2019}, we use the NN to predict the RK-stages for a fixed time step \timestep{} from which we then construct the state prediction \xpredictionhat{}:
\begin{align}
    \perceptron{0} &= [\xinitial, \inputvariable] \label{eq:RK_PINN_fixed_input}\\ 
    \outputNN &=  [{\RKstagehat{1\top}}, \dots, {\RKstagehat{s\top}} ]^\top \label{eq:RK_PINN_fixed_output}\\
    \xpredictionhat &=  \xinitial + \timestep \sum_{k=1}^s \RKbeta{k} \RKstagehat{k} \label{eq:RK_PINN_fixed_prediction}.
\end{align}
For all quantities that are based on a NN prediction, we will subsequently use the hat symbol $\hat{(\cdot)}$. In a supervised learning setting, we would be required to know the values for \RKstage{i} in order to adjust the NN's parameters accordingly. By making use of the fact that we obtain a ``correct" solution if \eqref{eq:RK_scheme} is satisfied, we can test how well the NN's predictions \RKstagehat{i} match the equations and define the vector \RKerror{k} for each RK-stage via
\begin{align}
    \RKerror{k}(\xinitial, \inputvariable) &= \RKstagehat{k} - \fvector \left(\tinitial + \RKgamma{k}\timestep, \xinitial + \timestep \sum_{l=1}^s \RKalpha{kl} \RKstagehat{l} ; \inputvariable \right). \label{eq:RKerror_fixed}
\end{align}
Based on \eqref{eq:RKerror_fixed}, we can evaluate the error element-wise \RKerrori{k} for each RK-stage and across $N$ collocation points (indexed by subscript $j$)
\begin{align}
    \lossRKstage &= \sum_{j=1}^{N} \RKerrori{k}(\xinitial_j, \inputvariable_j)^2\label{eq:RKerror_loss_fixed}
\end{align}
and formulate the training problem as
\begin{alignat}{2}
     \min_{\weights{i}, \biases{i}} \quad & \sum_{i, k} \lossRKstage && \label{eq:RK_PINN_fixed_loss}\\
     \text{s.t.}  \quad & \eqref{eq:NN_hidden_layers}, \eqref{eq:NN_output}, \eqref{eq:RK_PINN_fixed_input}, \eqref{eq:RK_PINN_fixed_output}. &&
\end{alignat}
Note, in the entire calculation, we are not required to know the ``true" values of \RKstage{k} or \xprediction{}, and hence, there is no need for running simulations to create a dataset of results for \RKstage{k} or \xprediction{}. Instead, we purely evaluate \RKerror{k} on a number of values for initial points \xinitial{} and inputs \inputvariable{}, i.e., the collocation points.

\subsection{Variable time step RK-PINN}
In the previous subsection, as well as in~\cite{Raissi:2019}, \timestep{} was a fixed value, and therefore, we would need to train different RK-PINNs if we wanted to incorporate different time steps. This paper extends the method by training a single NN, where \timestep{} is introduced as an input variable:
\begin{align}
    \perceptron{0} &= [\timestep, \xinitial, \inputvariable] \label{eq:RK_PINN_variable_input}.
\end{align}
The only other changes compared to the previous RK-PINN concern the error calculation of the RK-stages. We need to adjust \RKerror{k} to $\RKerror{k}(\timestep_j, \xinitial_j, \inputvariable_j)$ in \eqref{eq:RKerror_fixed} and \eqref{eq:RKerror_loss_fixed} since \timestep{} has become an input:
\begin{align}
    \lossRKstage &= \sum_{j=1}^{N} \RKerrori{k}(\timestep_j, \xinitial_j, \inputvariable_j)^2. \label{eq:RKerror_loss_variable}
\end{align}
We usually think of the implicit RK scheme as a discrete solver that is particularly useful for solving ODEs at a specific time step. By introducing the variable time steps, the RK-PINNs recover the characteristics of a continuous solution approximation such as with other PINNs. This in turn allows us to use the same regularisation which is commonly used with continuous time PINNs \cite{misyris_physics-informed_2020, stiasny_transient_2021}, in which we evaluate the consistency of the NN sensitivity with the governing differential equations. This error \physicsError{} is calculated by applying AD to \xpredictionhat{} and the comparing it with the state update in \eqref{eq:state_equation} when using the predicted state \xpredictionhat{} as an input:
\begin{align}
    \physicsError(\timestep, \xinitial, \inputvariable) &= \frac{\partial}{\partial \timestep} \xpredictionhat{} - \fvector \left(\tinitial + \timestep, \xpredictionhat{}; \inputvariable \right).
\end{align}
Analogous to \eqref{eq:RKerror_loss_variable}, we define a loss term \lossPredictionDiff{} across the training dataset
\begin{align}
    \lossPredictionDiff &= \sum_{j=1}^{N} \physicsError(\timestep_j, \xinitial_j, \inputvariable_j)^2.
\end{align}
This additional loss term still does not require any simulated data, but it places an additional regularisation on the neural network training procedure. The following optimisation problem describes the training setup in its final form:
\begin{alignat}{2}
     \min_{\weights{i}, \biases{i}} \quad & \sum_{i, k} \weightRKstage \lossRKstage + \sum_{i} \weightPredictionDiff \lossPredictionDiff && \label{eq:RK_PINN_variable_loss}\\
     \text{s.t.}  \quad & \eqref{eq:NN_hidden_layers}, \eqref{eq:NN_output},\eqref{eq:RK_PINN_fixed_output},  \eqref{eq:RK_PINN_fixed_prediction},  \eqref{eq:RK_PINN_variable_input}. &&
\end{alignat}
The coefficients \weightRKstage{} and \weightPredictionDiff{} serve to balance the influence of the different loss terms on the adjustment of the NN parameters \weights{i} and \biases{i}. This prevents a single loss term from dominating the optimisation problem, as this can lead to poor training characteristics.
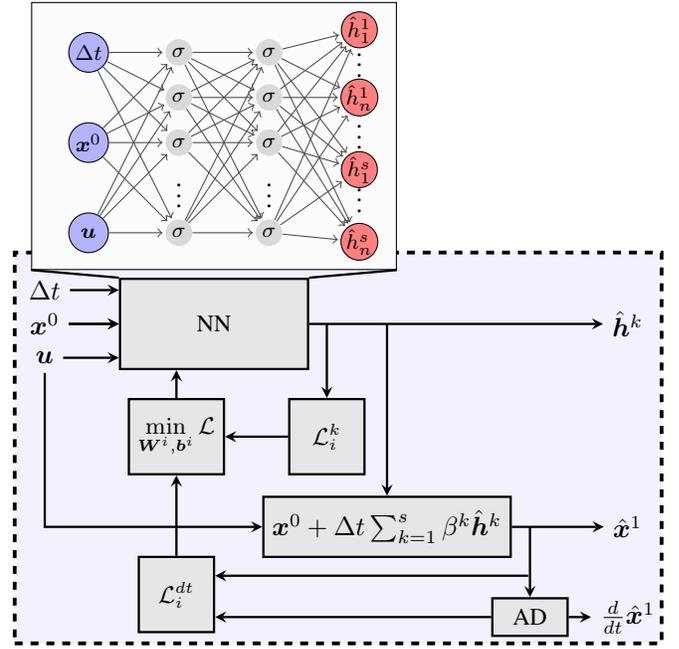
\begin{figure}
    \centering
    \input{figures/RK-PINN_architecture}
    \caption{Architecture of the RK-PINN.}
    \label{fig:RK-PINN_architecture}
\end{figure}

%% file: figures/RK-PINN_architecture.tex
\tikzstyle{BOXY} = [rectangle, rounded corners = 5, minimum width=10, minimum height=1.2cm,text centered, draw=black, fill=white,line width=0.3mm,font=\footnotesize, align=center]

\def\NNxshift{2.5cm}
\def\NNyshift{0.3cm}
\begin{tikzpicture}
    \node (NN_pure) [minimum width=8.6cm, minimum height=5.2cm,draw=black, dashed, line width=0.5mm, fill=blue!5] at (1.65cm, -1.65cm) {};
    \node (NN) [minimum width=2.5cm, minimum height=1.2cm,text centered, draw=black, fill=black!10,line width=0.3mm,font=\small, align=center] {NN};
    \node (input_t) [left of = NN, xshift = -0.5*\NNxshift, yshift = 1.5*\NNyshift] {\timestep{}};
    \node (input_x0) [left of = NN, xshift = -0.5*\NNxshift, yshift = 0*\NNyshift] {\xinitial{}};
    \node (input_u) [left of = NN, xshift = -0.5*\NNxshift, yshift = -1.5*\NNyshift] {\inputvariable{}};

    \node (output_x) [right of = NN, xshift = 1.8*\NNxshift, yshift = 0.0cm] {\RKstagehat{k}};
    \node (output_dx) [right of = NN, xshift = 1.8*\NNxshift, yshift = -2.7cm] {\xpredictionhat{}};
    \node (output_dxdt) [right of = NN, xshift = 1.8*\NNxshift, yshift = -3.9cm] {$\frac{d}{dt} \xpredictionhat{}$};
    
    \node (NNarchitecture) [rectangle, text centered,fill=black!1, align=center, draw=black, above of = NN, yshift = 1.5cm] {\input{figures/NN_architecture_RK_small}};
    
    \node (prediction) [rectangle, minimum width=0.8cm, minimum height=0.8cm, text centered, fill=black!10,line width=0.3mm, align=center, draw=black, left of = output_dx ,xshift = -2.2cm, yshift = 0.0cm] {$\xinitial + \timestep \sum_{k=1}^s \RKbeta{k} \RKstagehat{k}$};
    
    \node (optimize) [minimum width=1.0cm, minimum height=1.0cm, text centered, draw=black, fill=black!10,line width=0.3mm,font=\small, align=center, below of = NN, xshift = -0.5cm, yshift = -0.5cm] {$\min\limits_{\bm{W}^i, \bm{b}^i} \mathcal{L}$};
    
    \node (loss data) [minimum width=1.0cm, minimum height=1.0cm, text centered, draw=black, fill=black!10,line width=0.3mm,font=\small, align=center, right of = optimize, xshift = 1.0cm, yshift = -0.0cm] {\lossRKstage{}};
    
    \node (loss physics) [minimum width=1.0cm, minimum height=1.0cm, text centered, draw=black, fill=black!10,line width=0.3mm,font=\small, align=center, below of = optimize, xshift = 0.0cm, yshift = -1.1cm] {\lossPredictionDiff{}};
    
    \node (AD) [minimum width=1.0cm, minimum height=0.5cm, text centered, draw=black, fill=black!10,line width=0.3mm,font=\small, align=center, left of = output_dxdt, xshift = -0.3cm, yshift = 0.0cm] {AD};
    
    \node (AD helper) [above of = AD, xshift = 0.1cm, yshift = -0.45cm] {};
    

    \draw[black,->,>=stealth,line width=0.3mm] (input_t) -- (input_t -| NN.west);    
    \draw[black,->,>=stealth,line width=0.3mm] (input_u) -- (input_u -| NN.west);
    \draw[black,->,>=stealth,line width=0.3mm] (input_x0) -- (input_x0 -| NN.west);
    \draw[black,->,>=stealth,line width=0.3mm] (input_u) -- (input_u |- prediction.west) -- (prediction.west);
    \draw[black,->,>=stealth,line width=0.3mm] (input_x0) -- (input_x0 -| NN.west);
    \draw[black,->,>=stealth,line width=0.3mm] (NN.east) -- (output_x);   
    
    \draw[black,line width=0.3mm] (NN.north east) -- (NNarchitecture.south east); 
    \draw[black,line width=0.3mm] (NN.north west) -- (NNarchitecture.south west); 
    
    \draw[black,->,>=stealth,line width=0.3mm] (loss data.north |- NN.east) -- (loss data.north); 
    \draw[black,->,>=stealth,line width=0.3mm] (prediction.north |- NN.east) -- (prediction.north); 
    \draw[black,->,>=stealth,line width=0.3mm] (prediction.east) -- (output_dx); 
    \draw[black,->,>=stealth,line width=0.3mm] (optimize.north) -- (optimize.north |- NN.south);
    \draw[black,->,>=stealth,line width=0.3mm] (AD.east) -- (output_dxdt); 
    \draw[black,->,>=stealth,line width=0.3mm] (loss data.west) -- (optimize.east);
    
    \draw[black,->,>=stealth,line width=0.3mm] (prediction.east -| AD.north) -- (AD.north); 
    \draw[black,->,>=stealth,line width=0.3mm] (AD.west) -- (AD.west -| loss physics.east); 
    \draw[black,->,>=stealth,line width=0.3mm] (loss physics.north) -- (optimize.south); 
    \draw[black,->,>=stealth,line width=0.3mm] (AD helper.west) -- (AD helper.west -| loss physics.east);

\end{tikzpicture}

%% file: figures/NN_architecture_RK_small.tex
\pagestyle{empty}
\def\layersep{1.2cm}
\def\nodeinlayersep{0.6cm}
\begin{tikzpicture}[
   shorten >=1pt,->,
   draw=black!70,
    node distance=\layersep,
    every pin edge/.style={<-,shorten <=1pt},
    neuron/.style={circle,fill=black!25,minimum size=10pt,inner sep=0pt, font=\footnotesize},
    input neuron/.style={neuron, fill=blue!30, minimum size=15pt,draw=black},
    output neuron/.style={neuron, fill=red!50, minimum size=12pt,draw=black},
    hidden neuron/.style={neuron, fill=gray!30},
    operator neuron/.style={neuron, fill=yellow!50, minimum size=25pt,draw=black},
    summation neuron/.style={neuron, fill=gray!50, minimum size=18pt},
    parameter neuron/.style={neuron, fill=green!30, minimum size=40pt,draw=black},
    annot/.style={text width=4em, text centered}
]%
    \node[input neuron] (I-1) at (0,-1*\nodeinlayersep) {\timestep{}};%
    \node[input neuron] (I-2) at (0,-3*\nodeinlayersep) {\xinitial{}};%
    \node[input neuron] (I-3) at (0,-5*\nodeinlayersep) {\inputvariable{}};%
    \node[hidden neuron] (H1-1) at (1*\layersep,-1*\nodeinlayersep ) {$\sigma$};
    \node[hidden neuron] (H1-2) at (1*\layersep,-2*\nodeinlayersep ) {$\sigma$};
    \node[hidden neuron] (H1-3) at (1*\layersep,-3*\nodeinlayersep ) {$\sigma$};
    \node (H1-4) at (1*\layersep,-4*\nodeinlayersep ) {$\vdots$};
    \node[hidden neuron] (H1-5) at (1*\layersep,-5*\nodeinlayersep ) {$\sigma$};
    \node[hidden neuron] (H2-1) at (2*\layersep,-1*\nodeinlayersep ) {$\sigma$};
    \node[hidden neuron] (H2-2) at (2*\layersep,-2*\nodeinlayersep ) {$\sigma$};
    \node[hidden neuron] (H2-3) at (2*\layersep,-3*\nodeinlayersep ) {$\sigma$};
    \node (H2-4) at (2*\layersep,-4*\nodeinlayersep ) {$\vdots$};
    \node[hidden neuron] (H2-5) at (2*\layersep,-5*\nodeinlayersep ) {$\sigma$};
    \node[output neuron] (O-1) at (3*\layersep,-0.5*\nodeinlayersep) {$\hat{h}_1^1$};%
    \node (O-2) at (3*\layersep,-1.1*\nodeinlayersep) {$\vdots$};%
    \node[output neuron] (O-3) at (3*\layersep,-2.0*\nodeinlayersep) {$\hat{h}^1_n$};%
    \node (O-4) at (3*\layersep,-2.7*\nodeinlayersep) {$\vdots$};%
    \node[output neuron] (O-5) at (3*\layersep,-3.6*\nodeinlayersep) {$\hat{h}^s_1$};%
    \node (O-4) at (3*\layersep,-4.1*\nodeinlayersep) {$\vdots$};%
    \node[output neuron] (O-7) at (3*\layersep,-5.2*\nodeinlayersep) {$\hat{h}^s_n$};%
    \foreach \source in {1,2,3}
        \foreach \dest in {1,...,3,5} 
            \path (I-\source) edge (H1-\dest);%
    \foreach \source in {1,...,3,5}
        \foreach \dest in {1,...,3,5} 
            \path (H1-\source) edge (H2-\dest);%
    \foreach \source in {1,...,3,5}
        \foreach \dest in {1,3,5,7} 
            \path (H2-\source) edge (O-\dest);%
\end{tikzpicture}

%% file: sections/03_case_study.tex
\section{Case study}\label{sec:case_study}

We demonstrate the proposed methodology on a Single-Machine Infinite-Bus (SMIB) system, which we detail below. This section also specifies the NN parameters and the training setup.

\subsection{Single-Machine Infinite-Bus system}
\begin{figure}[ht]
    \centering
    \includegraphics[width=0.6\linewidth]{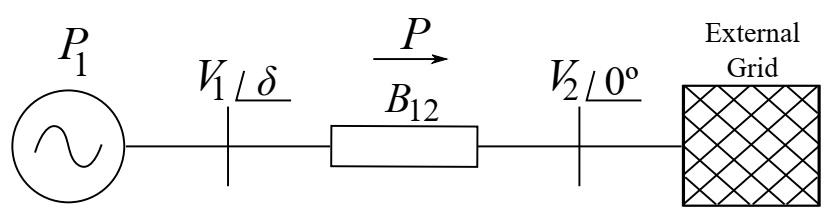}
    \caption{Single Machine Infinite Bus system}
    \label{fig:SMIB}
\end{figure}

The SMIB system shown in \cref{fig:SMIB} represents a second-order generator model connected to an external stiff grid. The voltage angle at the point of connection (Bus 2) is considered the reference angle; it is set constant and equal to $\delta_{ext} = \SI{0}{\radian}$. The state space equations \eqref{eq:SMIB} are time-invariant and assume a time-invariant active power production $P$. The system is furthermore parametrised by the machine's inertia constant $m = \SI{0.4}{\pu}$, a damping coefficient $d=\SI{0.15}{\pu}$, and the network parameters $B_{12}=\SI{0.2}{\pu}$ and voltages $V_1=V_2=\SI{1}{\pu}$.
\begin{align}\label{eq:SMIB}
   \frac{d}{dt} \begin{bmatrix} \delta \\ \omega \end{bmatrix} &= \begin{bmatrix} 0 & 1 \\ 0 & -\frac{d}{m} \end{bmatrix} \begin{bmatrix} \delta \\ \omega \end{bmatrix} + \begin{bmatrix} 0 \\ \frac{1}{m} (P - V_1 V_2 B_{12} \sin{\delta}) \end{bmatrix} 
\end{align}
We train the NN on the following domain ($\xinitial{} = \begin{bmatrix} \delta^0& \omega^0 \end{bmatrix}^\top$):
\begin{subequations}\label{eq:input_domain}
\begin{align}
    t &\in [0, 10] \si{\second}\label{eq:t_set}\\
    P &\in [0, 0.2] \si{\pu}\label{eq:P_set}\\
    \delta^0 &\in [-\frac{\pi}{2}, \frac{\pi}{2}] \si{\radian} \label{eq:d_set}\\
    \omega^0 &= 0.1 \si[per-mode = fraction]{\radian \per \second}.
\end{align}
\end{subequations}

\subsection{NN training}

For the training process, we use a single hidden layer $K = 1$ with 50 neurons and the $\tanh$ activation function. We test the approach for different numbers of RK-stages $s \in [4, 8, 16, 32]$. Our implementation utilises the TensorFlow framework \cite{abadi_tensorflow_nodate} and it is publicly available on github\footnote{Github repository: github.com/jbesty}. The training problem \eqref{eq:RK_PINN_variable_loss} is solved by using the stochastic gradient descent method Adam \cite{kingma_adam_2017} with a decaying learning rate of $0.05 \cdot 0.995^\frac{E}{100}$, where $E$ is the number of epochs. For the experiments, we first create a large database of points across the input domain with increments of \SI{0.1}{\second}, \SI{0.004}{\pu} and $\frac{\pi}{50}\,\si{\radian}$, corresponding to (\ref{eq:t_set}), (\ref{eq:P_set}), and (\ref{eq:d_set}), respectively. From this database, we randomly sample $N \in [50, 100, 200, 1000]$ collocation points, i.e., points that define \perceptron{0} but do not include the associated target values for \RKstage{k}. We train the models for 100'000 epochs, where an epoch refers to an optimisation step with respect to the loss function \lossGeneral{} in \eqref{eq:RK_PINN_variable_loss} across the collocation points. During the training we monitor the loss function \lossGeneral{} across another 1000 points, which serve as the validation set, and we use it for an early stopping of the training to prevent over-fitting. Lastly, we assess the accuracy of the model by evaluating the prediction error $e_{\delta}$:
\begin{align}
    \deltaerror = \left(\deltaprediction(\timestep, \delta^0, P) - \deltapredictionhat(\timestep, \delta^0, P) \right)^2
\end{align}
with \deltapredictionhat{} from $\xpredictionhat = \begin{bmatrix} \deltapredictionhat & \omegapredictionhat \end{bmatrix}^{\top}$. When presenting the results, we will use percentiles of the distribution of \deltaerror{} to describe its properties across a dataset. The $k$-th percentile refers to the value below which $k$ \% of the distribution lie.

%% file: sections/04_results.tex
\section{Results}\label{sec:results}
First, we will demonstrate the computational advantages of RK-PINNs for the evaluation of ODEs. Second, we show the high quality of the achieved accuracy for different RK-PINNs and point out the error characteristics of the method.

\subsection{Computational advantage in evaluation}

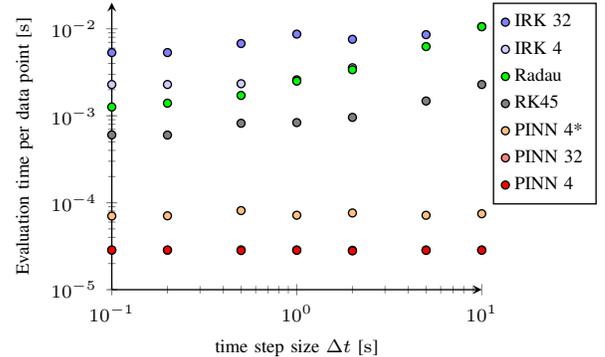
\begin{figure}[ht]
    \centering
    \input{figures/timing_plot}
    \caption{Evaluation time of different methods for increasing time step sizes. The legend indicates the method and the order. We note that the evaluation time difference between PINN 4 and PINN 32 is visually indistinguishable.}
    \label{fig:methods_timing}
\end{figure}

The primary advantage of using NNs for approximating the solution to ODEs lies in the speed of the evaluation. \Cref{fig:methods_timing} illustrates the enormous computational advantage of RK-PINNs. The most powerful feature of RK-PINNs is that the evaluation time is independent of the time step \timestep{} whereas the traditionally applied methods suffer from increasing computational effort with larger time steps, e.g., when we wish to determine the state of the system at $\Delta t=1s$ or $\Delta t=10s$. We show this trend for the implementation of full implicit RK (IRK) schemes with $s=4$ and $s=32$ RK-stages, respectively. For reference, we also use an implicit `Radau'-scheme and the explicit `RK-45'-scheme as provided in the \texttt{scipy.optimize} package. It is worth pointing out that increasing values for \timestep{} pose a challenge for lower order implicit RK schemes in terms of convergence. For example, IRK 4 in Fig.~\ref{fig:methods_timing} failed to converge beyond $\Delta t = 2s$. This motivates either moving to higher order and hence more expensive implicit RK schemes or using adaptive solution approaches. In contrast, RK-PINNs do not suffer from convergence issues; and as we will see in Section~\ref{sec:RK_stages_accuracy}, they achieve acceptable accuracy even with few RK-stages. As far as the computational cost of RK-PINNs is concerned, in contrast to the conventional approaches, this is primarily governed by the NN's size whereas the number of RK-stages plays a minor role. Since the involved calculations entail only a matrix multiplication and a non-linear function evaluation per NN layer, it is apparent that even a much larger RK-PINN would still retain a significant computational advantage over the other methods. To illustrate that, in Fig.~\ref{fig:methods_timing}, we compare a PINN with 3 layers and 500 neurons each, labelled PINN 4*, with our standard PINNs of 1 layer with 50 neurons, labelled PINN 4 and PINN 32.

It is important to note that \cref{fig:methods_timing} does not indicate anything about the accuracy of the RK-PINNs. For the other methods, the evaluation time reports the time required to reach a certain tolerance, in this case $10^{-13}$, for which each method will achieve a certain accuracy if it converges. Lowering numerical tolerance requirements can speed up the solution time at the risk of obtaining inaccurate solutions. For RK-PINNs, in contrast, the achieved accuracy is dependent on two aspects: i) the NN size as previously mentioned and ii) the training procedure. Hence, RK-PINNs of exactly the same size can have nearly identical evaluation times but achieve vastly different accuracy levels.

In essence, RK-PINNs, and in fact any similar NN architecture, will be much faster to evaluate than traditional ODE solvers; however, to exploit this advantage we have to show sufficient levels of accuracy. We explore this in the following subsections.

\subsection{Evaluating and interpreting accuracy}

\begin{figure}[ht]
    \centering
    \input{figures/trajectories}
    \caption{Prediction (coloured) and ground truth (black dashed) for trajectories}
    \label{fig:trajectories}
\end{figure}
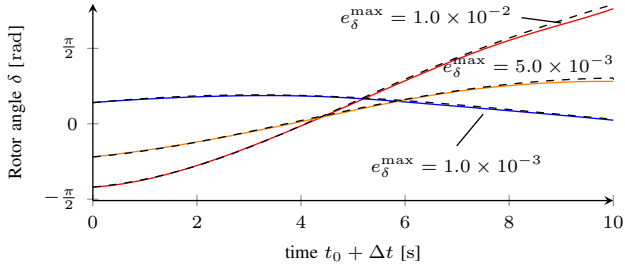

The main metric of interest is the achieved accuracy across the input domain defined in \eqref{eq:input_domain}. To give a first impression of the accuracy that we can accomplish with RK-PINNs, we consider three trajectories of $\delta$ in \cref{fig:trajectories}. The maximum error \deltaerror{} evaluates to $1.0\times 10^{-2}$ for the red, $5.0\times 10^{-3}$ for the yellow, and $1.0\times 10^{-3}$ for the blue trajectory.

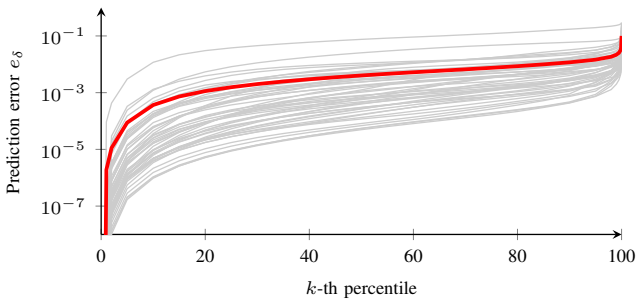
\begin{figure}[ht]
    \centering
    \input{figures/quantile_plot}
    \caption{Each grey line represents the error distribution on the test dataset of a trained NN. The red solid line is the mean of these error curves.}
    \label{fig:error_quantiles}
\end{figure}

Before comparing numeric metrics, it is worth taking a look at the distribution of the \deltaerror{} across the test dataset to give an understanding of how we arrive at the subsequent numbers. \Cref{fig:error_quantiles} depicts \deltaerror{} ordered by magnitude, i.e., the corresponding percentile is shown on the x-axis. Each grey line represents this error distribution for a trained RK-PINN. The different curves arise because of the random initialisation of the NN's weights and biases and the random sampling of the collocation points. We subsequently calculate the mean (shown in red in \cref{fig:error_quantiles}) and standard deviation of the percentile values to report the results in \cref{tbl:error_rk_stages} and \cref{tbl:error_dataset_size}.

\subsection{The effect of the number of RK-stages on accuracy}
\label{sec:RK_stages_accuracy}
\Cref{tbl:error_rk_stages} lists the results of the main experiment where we compare \deltaerror{} for RK-PINNs with different numbers of RK stages $s$. To be clear, a 4-stage RK-PINN outputs 4 RK state update prediction vectors $\RKstagehat{1},\dots,\RKstagehat{4}$, while a 32-stage RK-PINN outputs 32 such prediction vectors. The main insight is that for the given training setup more RK-stages $s$ improve both the mean and the standard deviation across the various training runs. This trend is consistently observable across all percentiles. This prediction accuracy must not be confused with the truncation error that reduces with higher order schemes; the present effect is the result of better approximations of \RKstage{k}. At this point, we point out once again that these results have purely been obtained by using the implicit RK-scheme and no simulation had to be performed in advance. The drawback from this are longer training times (in the order of tens of minutes), especially for higher-order systems.

\begin{table}[ht]
\renewcommand{\arraystretch}{1.3}
\caption{Mean and SD of the $k$-th percentile of \deltaerror{} for different implicit RK schemes implemented in RK-PINNs}
\label{tbl:error_rk_stages}
\centering
\begin{tabular}{lcccc}
    \toprule
    RK    & \multicolumn{4}{c}{$k$-th percentile} \\
    $s$  & 100  & 90       & 50   & 10  \\ 
        \midrule
4 &$1.33\pm0.96$ &$1.81\pm2.68$ &$6.29\pm12.1$ &$4.47\pm11.6$   \\
8 &$1.00\pm0.70$ &$1.17\pm2.57$ &$4.04\pm11.3$ &$3.67\pm16.8$      \\
16 & $0.65\pm0.45$ &$0.81\pm1.01$ &$2.53\pm4.03$ &$1.52\pm2.81$   \\
32  & $0.58\pm0.48$ &$0.35\pm0.63$ &$0.97\pm2.26$ &$0.60\pm2.31$   \\
     \midrule
      & $\times 10^{-1}$ & $\times 10^{-2}$ & $\times 10^{-3}$ & $\times 10^{-4}$ \\
\bottomrule
\end{tabular}
\end{table}

\subsection{The effect of the number of collocation points on accuracy}

\begin{figure}[ht]
    \centering
    \input{figures/error_characteristics}
    \caption{Distribution of the prediction error \deltaerror{} across the three input dimensions $(\timestep,P,\delta^0)$ for a network with $s=4$ and $N=50$ (left column) and $N = 1000$ (right column).}
    \label{fig:error_characteristics}
\end{figure}
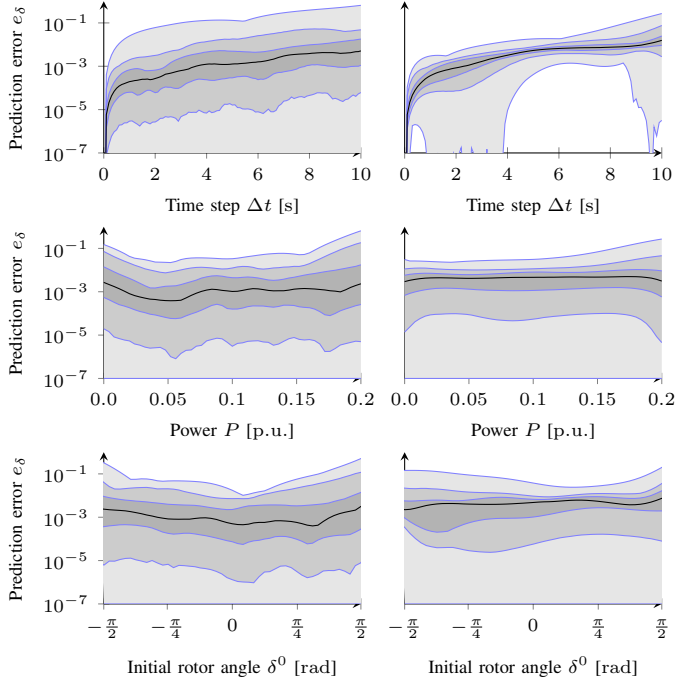

Beside the number of RK-stages, we investigate how the number of collocation points affects the accuracy. \Cref{tbl:error_dataset_size} presents the results and surprisingly, the outcome is not as clear as for the RK-stages. Whereas more points clearly lead to the smallest maximum errors (100th-percentile), fewer collocation points perform consistently better on lower percentiles. The answer lies in the error characteristics and the training process.
\begin{table}[ht]
\renewcommand{\arraystretch}{1.3}
\caption{Mean and SD of the $k$-th percentile  of \deltaerror{} for different numbers of collocation points $N$}
\label{tbl:error_dataset_size}
\centering
\begin{tabular}{lcccc}
    \toprule
       & \multicolumn{4}{c}{$k$-th percentile} \\
    $N$  & 100  & 90       & 50   & 10  \\  
        \midrule
50 &$14.2\pm15.7$ &$3.01\pm1.57$ &$2.31\pm1.27$ &$0.49\pm0.36$   \\
100 &$5.75\pm6.74$ &$1.09\pm0.73$ &$1.14\pm1.10$ &$0.29\pm0.45$      \\
200 & $2.21\pm1.76$ &$0.53\pm0.49$ &$1.03\pm1.89$ &$0.30\pm0.62$   \\
1000 &$1.33\pm0.96$ &$1.81\pm2.68$ &$6.29\pm12.1$ &$4.47\pm11.6$   \\
     \midrule
      & $\times 10^{-1}$ & $\times 10^{-2}$ & $\times 10^{-3}$ & $\times 10^{-4}$ \\
\bottomrule
\end{tabular}
\end{table}
If we consider the error across the test dataset for a single trained NN, we observe a fairly consistent distribution across each dimension in the input domain. This is shown in the subplots in \cref{fig:error_characteristics} for a case with only 50 collocation points on the left, and a case with 1000 collocation points on the right. The different shading indicates the bands in which 100\%, 90\%, 50\% of the error lies and the black line represents the median. These results suggest that, generally speaking, the RK-PINNs exhibit good generalisation abilities. However, more collocation points yield smoother and narrower yet worse approximations, except for the maximum error as it was indicated in \cref{tbl:error_dataset_size}. The less extreme errors can be explained by the fact that more collocation points allow better approximations at the input domain boundaries, a common phenomenon in NN training. \Cref{fig:error_characteristics} shows this behaviour, except for small time steps where all \RKstagehat{k} should tend to $\bm{0}$, hence, potential approximation errors are usually small. The reason for the slightly worse overall predictions with $N=1000$ remains to be investigated, but it seems that more collocation points complicate the training process. Eventually, a trade-off between potentially longer training times and the danger of larger errors has to be found. 

%% file: figures/timing_plot.tex
\pagestyle{empty}

\def\subplotWidth{6.5cm}
\def\horizontalDistance{5cm}
\def\subplotHeight{5.4cm}
\def\heightOffset{0.5cm}


\begin{tikzpicture}[every node/.style={font=\scriptsize},
    x=0cm,
    y=1cm]%
    \begin{loglogaxis}[
    xlabel near ticks, 
    ylabel near ticks, 
    width=\subplotWidth, 
    height=\subplotHeight,
    yshift=0,
    xshift=0,
    line width=0.5,
    ylabel={Evaluation time per data point [\si{\second}]},
    ymajorticks=true,
    axis x line=bottom,
    axis y line=left,
    xlabel={time step size \timestep{} [\si{\second}]},
    ymin = 0.00001,
    ymax = 0.02,
    xmin = 0.1,
    xmax = 10,
    legend style={font=\footnotesize}, 
    legend pos=outer north east,
    legend cell align={left},
    legend columns=1,
    ]%
    \addplot[only marks, mark size=1.5pt, fill=blue!50] table[x index=0, y index=2] {data_evaluation_time.dat};
    \addlegendentry{IRK 32};
    \addplot[only marks, mark size=1.5pt, fill=blue!20] table[x index=0, y index=1] {data_evaluation_time.dat};
    \addlegendentry{IRK 4};
    \addplot[only marks, mark size=1.5pt, fill=green] table[x index=0, y index=3] {data_evaluation_time.dat};
    \addlegendentry{Radau};
    \addplot[only marks, mark size=1.5pt, fill=gray] table[x index=0, y index=4] {data_evaluation_time.dat};
    \addlegendentry{RK45};
    \addplot[only marks, mark size=1.5pt, fill=orange!50] table[x index=0, y index=7] {data_evaluation_time.dat};
    \addlegendentry{PINN 4*};
    \addplot[only marks, mark size=1.5pt, fill=red!50] table[x index=0, y index=6] {data_evaluation_time.dat};
    \addlegendentry{PINN 32};
    \addplot[only marks, mark size=1.5pt, fill=red] table[x index=0, y index=5] {data_evaluation_time.dat};
    \addlegendentry{PINN 4};

    \end{loglogaxis}
\end{tikzpicture}
    

%% file: figures/trajectories.tex
\pagestyle{empty}

\def\subplotWidth{8.5cm}
\def\horizontalDistance{5cm}
\def\subplotHeight{4.2cm}
\def\heightOffset{0.5cm}

\begin{tikzpicture}[every node/.style={font=\scriptsize},
    x=0cm,
    y=1cm]%
    \begin{axis}[
    xlabel near ticks, 
    ylabel near ticks, 
    width=\subplotWidth, 
    height=\subplotHeight,
    yshift=0,
    xshift=0,
    line width=0.5,
    ylabel={Rotor angle $\delta$ [\si{\radian}]},
    ymajorticks=true,
    axis x line=bottom,
    axis y line=left,
    xlabel={time $\tinitial +\timestep{}$ [\si{\second}]},
    ymin = -1.6,
    ymax = 2.5,
    xmin = 0.0,
    xmax = 10,
    ytick={-1.571, 0, 1.571},
    yticklabels={$-\frac{\pi}{2}$, $0$, $\frac{\pi}{2}$},
    ]%
    \addplot[color=red] table[x index=1, y index=7] {trajectories.dat};
    \addplot[color=blue] table[x index=1, y index=5] {trajectories.dat};
    \addplot[color=orange] table[x index=1, y index=6] {trajectories.dat};
    \addplot[color=black, dashed] table[x index=1, y index=2] {trajectories.dat};
    \addplot[color=black, dashed] table[x index=1, y index=3] {trajectories.dat};
    \addplot[color=black, dashed] table[x index=1, y index=4] {trajectories.dat};
    \draw [-] (axis cs:9,2.1)-- +(-15pt,2pt) node[left] {$\deltaerror^{\max} = 1.0\times 10^{-2}$};
    \draw [-] (axis cs:7.5,0.2)-- +(-10pt,-12pt) node[below] {$\deltaerror^{\max} = 1.0\times 10^{-3}$};
    \draw [-] (axis cs:10.0,0.92)-- +(4pt,5pt) node[left, align=center] {$\deltaerror^{\max} = 5.0\times 10^{-3}$};
    \end{axis}

\end{tikzpicture}
    

%% file: figures/quantile_plot.tex
\pagestyle{empty}

\def\subplotWidth{8.5cm}
\def\horizontalDistance{5cm}
\def\subplotHeight{4.6cm}
\def\heightOffset{0.5cm}

\begin{tikzpicture}[every node/.style={font=\scriptsize},
    x=0cm,
    y=1cm]%
    \begin{semilogyaxis}[
    xlabel near ticks, 
    ylabel near ticks, 
    width=\subplotWidth, 
    height=\subplotHeight,
    yshift=0,
    xshift=0,
    line width=0.5,
    ylabel={Prediction error \deltaerror{}},
    ymajorticks=true,
    axis x line=bottom,
    axis y line=left,
    xlabel={$k$-th percentile},
    ymin = 0.00000001,
    ymax = 1.0,
    xmin = 0.0,
    xmax = 1.0,
    ytick={0.1, 0.001, 0.00001, 0.0000001},
    xtick={0, 0.2, 0.4, 0.6, 0.8, 1.0},
    xticklabels={0, 20, 40, 60, 80, 100},
    ]%
    \foreach \n in {1,...,49} {
        \addplot[draw=black!20] table[y index=\n] {data_quantile_plot.dat};
	}
	
	\addplot[draw=red, line width=0.5mm] table[x index=0, y index=1] {data_quantile_plot_mean.dat};
    \end{semilogyaxis}
\end{tikzpicture}
    

%% file: figures/error_characteristics.tex
\pagestyle{empty}

\def\subplotWidth{5cm}
\def\horizontalDistance{2cm}
\def\subplotHeight{3.6cm}
\def\heightOffset{0.6cm}
\def\pathcolor{blue!50}

\begin{tikzpicture}[every node/.style={font=\scriptsize},
    xshift=0,
    yshift=2.3*\subplotHeight]%
    \begin{semilogyaxis}[
    xlabel near ticks, 
    ylabel near ticks, 
    width=\subplotWidth, 
    height=\subplotHeight,
    yshift=2*\subplotHeight-2*\heightOffset,
    xshift=-\horizontalDistance,
    line width=0.5,
    ylabel={Prediction error \deltaerror{}},
    ymajorticks=true,
    axis x line=bottom,
    axis y line=left,
    xlabel={Time step \timestep{} [\si{\second}]},
    ytick={0.1, 0.001, 0.00001, 0.0000001},
    ymin = 0.0000001,
    ymax = 1,
    xmin = 0.0,
    xmax = 10,
    ]%
    \addplot[color=\pathcolor, name path = q000, domain = 0:10] table[x index=1, y index=3] {data_error_time.dat};
    \addplot[color=\pathcolor, name path = q100, domain = 0:10] table[x index=1, y index=9] {data_error_time.dat};
    \addplot[gray!20] fill between [of = q000 and q100];
    \addplot[color=\pathcolor, name path = q005, domain = 0:10] table[x index=1, y index=4] {data_error_time.dat};
    \addplot[color=\pathcolor, name path = q095, domain = 0:10] table[x index=1, y index=8] {data_error_time.dat};
    \addplot[gray!40] fill between [of = q005 and q095];
    \addplot[color=\pathcolor, name path = q025, domain = 0:10] table[x index=1, y index=5] {data_error_time.dat};
    \addplot[color=\pathcolor, name path = q075, domain = 0:10] table[x index=1, y index=7] {data_error_time.dat};
    \addplot[gray!60] fill between [of = q025 and q075];
    \addplot[color=black] table[x index=1, y index=6] {data_error_time.dat};
    \end{semilogyaxis}
    
    \begin{semilogyaxis}[
    xlabel near ticks, 
    ylabel near ticks, 
    width=\subplotWidth, 
    height=\subplotHeight,
    yshift=1*\subplotHeight-\heightOffset,
    xshift=-\horizontalDistance,
    line width=0.5,
    ylabel={Prediction error \deltaerror{}},
    ymajorticks=true,
    axis x line=bottom,
    axis y line=left,
    xlabel={Power $P$ [\si{\pu}]},
    xtick={0, 0.05, 0.1, 0.15, 0.2},
    xticklabels={$0.0$, $0.05$, $0.1$, $0.15$, $0.2$},
    ytick={0.1, 0.001, 0.00001, 0.0000001},
    ymin = 0.0000001,
    ymax = 1.0,
    xmin = 0.0,
    xmax = 0.2,
    ]%
    \addplot[color=\pathcolor, name path = q000, domain = 0:0.2, samples = 10] {0.0000001};
    \addplot[color=\pathcolor, name path = q100, domain = 0:0.2] table[x index=1, y index=9] {data_error_power.dat};
    \addplot[gray!20] fill between [of = q000 and q100];
    \addplot[color=\pathcolor, name path = q005, domain = 0:0.2] table[x index=1, y index=4] {data_error_power.dat};
    \addplot[color=\pathcolor, name path = q095, domain = 0:0.2] table[x index=1, y index=8] {data_error_power.dat};
    \addplot[gray!40] fill between [of = q005 and q095];
    \addplot[color=\pathcolor, name path = q025, domain = 0:0.2] table[x index=1, y index=5] {data_error_power.dat};
    \addplot[color=\pathcolor, name path = q075, domain = 0:0.2] table[x index=1, y index=7] {data_error_power.dat};
    \addplot[gray!60] fill between [of = q025 and q075];
    \addplot[color=black] table[x index=1, y index=6] {data_error_power.dat};
    \end{semilogyaxis}
    
    \begin{semilogyaxis}[
    xlabel near ticks, 
    ylabel near ticks, 
    width=\subplotWidth, 
    height=\subplotHeight,
    yshift=0*\subplotHeight,
    xshift=-\horizontalDistance,
    line width=0.5,
    ylabel={Prediction error \deltaerror{}},
    ymajorticks=true,
    axis x line=bottom,
    axis y line=left,
    xlabel={Initial rotor angle $\delta^0$ [\si{\radian}]},
    xtick={-1.51, -0.764, 0, 0.764, 1.51},
    xticklabels={$-\frac{\pi}{2}$,$-\frac{\pi}{4}$, $0$, $\frac{\pi}{4}$, $\frac{\pi}{2}$},
    ytick={0.1, 0.001, 0.00001, 0.0000001},
    ymin = 0.0000001,
    ymax = 1.0,
    xmin = -1.51,
    xmax = 1.51,
    ]%
    \addplot[color=\pathcolor, name path = q000, domain = -1.51:1.51, samples = 10] {0.0000001};
    \addplot[color=\pathcolor, name path = q100, domain = -1.51:1.51] table[x index=1, y index=9] {data_error_delta_initial.dat};
    \addplot[gray!20] fill between [of = q000 and q100];
    \addplot[color=\pathcolor, name path = q005, domain = -1.51:1.51] table[x index=1, y index=4] {data_error_delta_initial.dat};
    \addplot[color=\pathcolor, name path = q095, domain = -1.51:1.51] table[x index=1, y index=8] {data_error_delta_initial.dat};
    \addplot[gray!40] fill between [of = q005 and q095];
    \addplot[color=\pathcolor, name path = q025, domain = -1.51:1.51] table[x index=1, y index=5] {data_error_delta_initial.dat};
    \addplot[color=\pathcolor, name path = q075, domain = -1.51:1.51] table[x index=1, y index=7] {data_error_delta_initial.dat};
    \addplot[gray!60] fill between [of = q025 and q075];
    \addplot[color=black] table[x index=1, y index=6] {data_error_delta_initial.dat};
    \end{semilogyaxis}
    
    \begin{semilogyaxis}[
    xlabel near ticks, 
    ylabel near ticks, 
    width=\subplotWidth, 
    height=\subplotHeight,
    yshift=2*\subplotHeight-2*\heightOffset,
    xshift=\horizontalDistance,
    line width=0.5,
    ymajorticks=false,
    axis x line=bottom,
    axis y line=left,
    xlabel={Time step \timestep{} [\si{\second}]},
    ymin = 0.0000001,
    ymax = 1,
    xmin = 0.0,
    xmax = 10,
    legend style={font=\footnotesize, 
    at={(1.1,0.55)}},
    legend columns=1,
    ]%
    \addplot[color=\pathcolor, name path = q000, domain = 0:10] table[x index=1, y index=3] {data_error_time_1000.dat};
    \addplot[color=\pathcolor, name path = q100, domain = 0:10] table[x index=1, y index=9] {data_error_time_1000.dat};
    \addplot[gray!20] fill between [of = q000 and q100];
    \addplot[color=\pathcolor, name path = q005, domain = 0:10] table[x index=1, y index=4] {data_error_time_1000.dat};
    \addplot[color=\pathcolor, name path = q095, domain = 0:10] table[x index=1, y index=8] {data_error_time_1000.dat};
    \addplot[gray!40] fill between [of = q005 and q095];
    \addplot[color=\pathcolor, name path = q025, domain = 0:10] table[x index=1, y index=5] {data_error_time_1000.dat};
    \addplot[color=\pathcolor, name path = q075, domain = 0:10] table[x index=1, y index=7] {data_error_time_1000.dat};
    \addplot[gray!60] fill between [of = q025 and q075];
    \addplot[color=black] table[x index=1, y index=6] {data_error_time_1000.dat};
    \end{semilogyaxis}
    
    
    \begin{semilogyaxis}[
    xlabel near ticks, 
    ylabel near ticks, 
    width=\subplotWidth, 
    height=\subplotHeight,
    yshift=1*\subplotHeight-\heightOffset,
    xshift=\horizontalDistance,
    line width=0.5,
    ymajorticks=false,
    axis x line=bottom,
    axis y line=left,
    xlabel={Power $P$ [\si{\pu}]},
    xtick={0, 0.05, 0.1, 0.15, 0.2},
    xticklabels={$0.0$, $0.05$, $0.1$, $0.15$, $0.2$},
    ymin = 0.0000001,
    ymax = 1.0,
    xmin = 0,
    xmax = 0.2,
    ]%
    \addplot[color=\pathcolor, name path = q000, domain = 0:0.2, samples = 10] {0.0000001};
    \addplot[color=\pathcolor, name path = q100, domain = 0:0.2] table[x index=1, y index=9] {data_error_power_1000.dat};
    \addplot[gray!20] fill between [of = q000 and q100];
    \addplot[color=\pathcolor, name path = q005, domain = 0:0.2] table[x index=1, y index=4] {data_error_power_1000.dat};
    \addplot[color=\pathcolor, name path = q095, domain = 0:0.2] table[x index=1, y index=8] {data_error_power_1000.dat};
    \addplot[gray!40] fill between [of = q005 and q095];
    \addplot[color=\pathcolor, name path = q025, domain = 0:0.2] table[x index=1, y index=5] {data_error_power_1000.dat};
    \addplot[color=\pathcolor, name path = q075, domain = 0:0.2] table[x index=1, y index=7] {data_error_power_1000.dat};
    \addplot[gray!60] fill between [of = q025 and q075];
    \addplot[color=black] table[x index=1, y index=6] {data_error_power_1000.dat};
    \end{semilogyaxis}
    
    \begin{semilogyaxis}[
    xlabel near ticks, 
    ylabel near ticks, 
    width=\subplotWidth, 
    height=\subplotHeight,
    yshift=0*\subplotHeight,
    xshift=\horizontalDistance,
    line width=0.5,
    ymajorticks=false,
    axis x line=bottom,
    axis y line=left,
    xlabel={Initial rotor angle $\delta^0$ [\si{\radian}]},
    xtick={-1.51, -0.764, 0, 0.764, 1.51},
    xticklabels={$-\frac{\pi}{2}$,$-\frac{\pi}{4}$, $0$, $\frac{\pi}{4}$, $\frac{\pi}{2}$},
    ymin = 0.0000001,
    ymax = 1.0,
    xmin = -1.51,
    xmax = 1.51,
    ]%
    \addplot[color=\pathcolor, name path = q000, domain = -1.51:1.51, samples = 10] {0.0000001};
    \addplot[color=\pathcolor, name path = q100, domain = -1.51:1.51] table[x index=1, y index=9] {data_error_delta_initial_1000.dat};
    \addplot[gray!20] fill between [of = q000 and q100];
    \addplot[color=\pathcolor, name path = q005, domain = -1.51:1.51] table[x index=1, y index=4] {data_error_delta_initial_1000.dat};
    \addplot[color=\pathcolor, name path = q095, domain = -1.51:1.51] table[x index=1, y index=8] {data_error_delta_initial_1000.dat};
    \addplot[gray!40] fill between [of = q005 and q095];
    \addplot[color=\pathcolor, name path = q025, domain = -1.51:1.51] table[x index=1, y index=5] {data_error_delta_initial_1000.dat};
    \addplot[color=\pathcolor, name path = q075, domain = -1.51:1.51] table[x index=1, y index=7] {data_error_delta_initial_1000.dat};
    \addplot[gray!60] fill between [of = q025 and q075];
    \addplot[color=black] table[x index=1, y index=6] {data_error_delta_initial_1000.dat};
    \end{semilogyaxis}
\end{tikzpicture}

%% file: sections/05_conclusion.tex
\section{Conclusion}\label{sec:conclusion}

In this work, we have introduced an approach, known as RK-PINNs, for solving parameterised power system ODEs. As shown by our results, RK-PINNs are capable of reducing the computation time required by conventional solvers by up to two orders of magnitude. Unlike many conventional MOR methods, RK-PINNs can provide an arbitrary degree of approximation accuracy of any continuous function; thus, their applicability is not limited to any narrow range of problems. Furthermore, our results indicate that RK-PINNs can be trained on, and successfully predict, low order RK integration schemes which conventional numerical solvers fail to solve entirely (given a sufficiently large time step). Given their excellent predictive accuracy, as showcased in this paper, RK-PINNs can thus be used for extremely fast screening of a very large number of potential power system contingencies. Future work will focus on (i) reducing the training time required by RK-PINNs, (ii) developing enhanced approaches for selecting optimally representative collocation points, and (iii) exploring training optimisation approaches which will scale favorably with larger model system models.

